\documentclass[12pt,preprint]{aastex}

\newcommand{\ba}{\begin{eqnarray}}
\newcommand{\ea}{\end{eqnarray}}
\newcommand{\be}{\begin{equation}}
\newcommand{\ee}{\end{equation}}
\def\eps{\epsilon}
\def\veps{\varepsilon}

\begin{document}

\title{Prompt TeV Emission from Cosmic Rays Accelerated by Gamma Ray
  Bursts Interacting with Surrounding Stellar Wind}
 
\author{Soebur Razzaque\altaffilmark{1,2}, Olga Mena\altaffilmark{3},
  and Charles D. Dermer\altaffilmark{1}}
 
\altaffiltext{1}{Space Science Division, U.S. Naval Research
  Laboratory, Washington, DC 20375, USA. e-mail:
  srazzaque@ssd5.nrl.navy.mil }
 
\altaffiltext{2} {National Research Council Research Associate}
 
\altaffiltext{3} {IFIC (CSIC -- Univ. de Valencia), Spain}
 
\begin{abstract} Protons accelerated in the internal shocks of a
long duration gamma ray burst can escape the fireball as cosmic rays
by converting to neutrons.  Hadronic interactions of these neutrons
inside a stellar wind bubble created by the progenitor star will
produce TeV gamma rays via neutral meson decay and synchrotron
radiation by charged pion-decay electrons in the wind magnetic
field. Such gamma rays should be observable from nearby gamma ray
bursts by currently running and upcoming ground-based
detectors. \end{abstract}


\section{Introduction}

TeV $\gamma$-rays have not been yet convincingly detected from a gamma
ray burst (GRB).  The data from GRB 970417a reported by the Milagrito
water Cherenkov detector~\citep{milagrito00} and from GRB 971110
reported by the GRAND air shower array~\citep{grand03} lack one of the
most crucial pieces of information about those GRBs, namely their
redshifts. A redshift is necessary to estimate the burst energetics
and to properly take into account attenuation of TeV $\gamma$-rays in
the extragalactic background radiation (EBL) fields.  Moreover, their
detection is at the $\approx 3\sigma$ level and hence statistically
not very significant.  A detection at higher significance level has
been reported by the TIBET air shower array by stacking data for a
large number of GRB time windows~\citep{tibet01}.  Currently, several
imaging air Cherenkov telescopes (IACTs) such as MAGIC and VERITAS are
capable of slewing to the GRB direction in the sky prompted by burst
alert network.  The successor of the recently decomissioned Milagro
detector, namely the High Altitude Water Cherenkov detector
HAWC\footnote{http://umdgrb.umd.edu/hawc/}, has a high duty cycle and
is particularly suitable to detect $\gtrsim$TeV $\gamma$-rays from
GRB.  Their observations have provided upper limits on $\gamma$-ray
fluence from several GRBs~\citep{milagro05,magic06}.  More powerful
detectors covering a wider energy range such as AGIS~\citep{agis07}
and CTA\footnote{http://www.mpi-hd.mpg.de/hfm/CTA/} are being planned.

Theoretical models do not predict TeV $\gamma$-rays from the internal
shocks of a GRB because of a high opacity to $e^\pm$ pair production
with observed keV--MeV energy $\gamma$-rays unless the Lorentz factor
of the relativistic bulk motion is very large~\citep{rmz04}.  In the
external shock with larger shock radii, TeV $\gamma$-rays formed by
Compton scattering of shock-accelerated electrons may avoid $e^\pm$
pair production ~\citep{dermer00,wdl01,zm01}. Photohadronic cascades
induced by shock-accelerated protons~\citep{bd98} requires intense
internal soft photon fields that will strongly attenuate TeV
$\gamma$-rays in the EBL.

In this Letter we propose a hadronic mechanism to produce TeV
$\gamma$-rays at the same time as the prompt keV--MeV emission.  If
protons are accelerated in the internal shocks of a GRB~\citep{wax95},
they are expected to interact with observed keV--MeV photons to
produce neutrons ($p\gamma \to n\pi^+$) which may escape the shock
region as cosmic-rays~\citep{wb97,rm98,da03}.  A fraction of these
cosmic rays will interact with particles in a surrounding dense
stellar wind as the progenitor star is expected to undergo substantial
mass loss before explosion~\citep{cl99}.  Neutral pions from secondary
nuclear production promptly decay to produce very high energy
$\gamma$-rays, while charged pions decay to produce electrons which
emit synchrotron radiation in the magnetic field of the stellar wind.
These $\gamma$-rays at TeV energy could be detected if they avoid
substantial absorption in the source environment as well as

We find that TeV $\gamma$-rays formed by nuclear interactions of
escaping neutrons with stellar wind particles may be detected from a
nearby GRB by current and upcoming $\gamma$-ray Cherenkov telescopes.

\section{GRB internal shocks and cosmic-ray escape}

The GRB internal shocks take place over a wide range of fireball
radii, depending on the $\gamma$-ray variability time scale $t_v \sim
10^{-3}$~s and the Lorentz factor of the bulk outflow $\Gamma_{b}
\gtrsim 10^{2.5}\Gamma_{b,2.5}$ for typical long GRBs. For observed
non-thermal emission, the radii where the internal shocks occur need
to be larger than the jet photospheric radius $r_{\rm ph} \cong
(\sigma_{T} L_{\gamma,\rm iso})/ (4\pi \eps_{e} \Gamma_b^3 m_p c^3)
\approx 3.7\times 10^{11} L_{\gamma,51} \eps_{e,-1}^{-1}
\Gamma_{b,2.5}^{-3}$~cm, at which the fireball becomes optically thin
to Thomson scattering.  Here we use an isotropic-equivalent
$\gamma$-ray luminosity of $L_{\gamma,\rm iso} = 10^{51}L_{\gamma,51}$
ergs s$^{-1}$ and a kinetic luminosity of $L_{\rm k,iso} =
L_{\gamma,\rm iso} \eps_e^{-1}$, where $\eps_e = 0.1\eps_{e,-1}$ is
the fraction of kinetic energy converted to $\gamma$-rays (assuming a
{\em fast-cooling} scenario).  With these parameters we calculate a
pre-shock electron and baryon number density of $n'_e \cong n'_p \cong
L_{\gamma,\rm iso}/(4\pi \eps_{e} r_{\rm sh}^2 \Gamma_b^2 m_p c^3)$ in
the comoving frame.  We denote the variables in the comoving plasma
(observer's) frame with (without) primes.

For our modeling purpose, we assume a shock radius of $r_{\rm sh} =
10^{12}r_{\rm sh,12}$~cm. The turbulent magnetic field strength in the
shock region, assuming that the magnetic energy density $B^{'2}_{\rm
sh}/8\pi$ is a fraction $\eps_B = 0.1\eps_{B,-1}$ of the fireball's
kinetic energy density $n'_p m_pc^2$, is
\be
B'_{\rm sh} 
\approx 8.2\times 10^5~ \eps_{B,-1}^{1/2} L_{\gamma,51}^{1/2} 
\eps_{e,-1}^{-1/2} r_{\rm sh,12}^{-1} \Gamma_{b,2.5}^{-1}~ {\rm G}.
\label{B_field}
\ee
The protons and electrons are assumed to be accelerated via a Fermi
mechanism by this magnetic field.

Characteristic synchrotron photons radiated by the population of
electrons with a minimum Lorentz factor $\gamma'_{e,\rm min} \simeq
\eps_e (m_p/m_e)(\Gamma_{\rm rel}-1)$ is one of the leading models to
produce observed $\gamma$-rays. With the parameters adopted here and a
relative Lorentz factor between two colliding shells $\Gamma_{\rm
  rel}\approx 3$, the observed characteristic synchrotron photon
energy is
\ba
\veps_{m} &\cong &
(3/2) (B'_{\rm sh}/B_Q) \gamma_{e,\rm min}^{'2} \Gamma_b m_ec^2 /(1+z)
\nonumber \\ &\approx & 600 (1+z)^{-1} 
\eps_{e,-1}^{3/2} \eps_{B,-1}^{1/2} 
L_{\gamma,51}^{1/2} r_{\rm sh, 12}^{-1}~ {\rm keV}.
\label{nu_m}
\ea
Here $B_Q = m_e^2 c^3 /q\hbar \approx 4.414\times 10^{13}$~G. The
observed photon energy at the peak of the $\veps F_\veps$ spectrum,
$\veps_{\rm pk}$, varies from burst to burst, however, there exist
several phenomenological relations connecting the peak photon energy
to other burst parameters~\citep[e.g.,][]{amati,ghirlanda,will08}.
Here we adopt a relation between the peak $\gamma$-ray luminosity and
a characteristic photon energy, which in turn is related to
$\veps_{\rm pk}$ as found by~\citet{will08}.  We rewrite this
relationship as
\be 
\veps_{\rm pk} \approx 650 (1+z)^{-1} L_{\gamma,51}^{0.27} ~{\rm keV}.
\label{Lpk-Epk-relation}
\ee
Note that this is close to the value of the synchrotron photon energy
in equation~(\ref{nu_m}).

Following the phenomenological broken power-law fits, we write the
comoving photon spectrum as
\ba 
n'_\gamma (\veps') &\cong & n'_{\gamma,\rm pk} \Gamma_b
/[\veps_{\rm pk} (1+z)] \nonumber \\
&& \times \cases{
(\veps'_{\rm sa}/\veps'_{\rm pk})^{-\alpha} 
(\veps'/\veps'_{\rm sa})^{3/2} ;~ \veps' < \veps'_{\rm sa} \cr
(\veps'/\veps'_{\rm pk})^{-\alpha} 
;~ \veps'_{\rm sa} \le \veps' \le \veps'_{\rm pk} \cr
(\veps'/\veps'_{\rm pk})^{-\beta}
;~ \veps'_{\rm max} > \veps' > \veps'_{\rm pk} ~,
}
\label{soft_photon_spectrum}
\ea
where $(\veps'_{\rm sa},~\veps'_{\rm max}) = (10^{-2.5},~10^6)$~keV
are respectively the synchrotron self-absorption and maximum photon
energies. The fitted values for the power-law indices are
$(\alpha,~\beta) = (1,~2.3)$. We calculate the peak photon number
density $L_{\gamma,\rm iso}/(12\pi r_{\rm sh}^2 c \Gamma_{b}
\veps_{\rm pk})$, including a bolometric factor of $\sim3$ and using
equation~(\ref{Lpk-Epk-relation}), as
\be
n'_{\gamma,\rm pk} \approx 2.7\times 10^{18} 
(1+z) L_{\gamma,51}^{0.73} \Gamma_{b,2.5}^{-1} r_{\rm sh,12}^{-2}
~{\rm cm}^{-3}.
\label{pk_photon_density}
\ee

The energy gained by the protons is proportional to the time
$t'_{p,\rm acc} \simeq \phi E'_p/qB'_{\rm sh}c$, they spend in the
shock region.  The maximum energy is typically obtained by requiring
that this time with $\phi\gtrsim 1$ to be equivalent to the smaller of
the fireball expansion or dynamic time $t'_{\rm dyn} \simeq r_{\rm
  sh}/2c\Gamma_b$ and the energy loss time scale $t'_{p,\rm loss}$.
The synchroton energy loss time scale $t'_{p,\rm syn} \cong 6\pi m_p^4
c^3/(\sigma_{T} m_e^2 E'_p B_{\rm sh}^{'2})$ is typically the shortest
for internal shocks. A maximum cosmic ray proton energy can thus be
obtained as
\be
E_{p,\rm max} 
\approx 7\times 10^{10}
\eps_{e,-1}^{1/4} \Gamma_{b,2.5}^{3/2} r_{\rm sh,12}^{1/2}
\eps_{B,-1}^{-1/4} L_{\gamma,51}^{-1/4} ~{\rm GeV},
\label{max_CR_energy}
\ee
for $t'_{\rm acc} = t'_{\rm syn}$ .  The differential spectrum (e.g.,
in units of cm$^{-2}$~s$^{-1}$~GeV$^{-1}$) of cosmic ray protons, if
they could escape freely from the fireball at a luminosity distance
$d_L$, may be written as
\be
J_p (E_p) \cong  L_{\gamma,\rm iso}/[ 4\pi d_{L}^2 \eps_e E_p^{2} 
~{\rm ln} (E_{p,\rm max}/\Gamma_b m_p c^2)],
\label{CR_flux}
\ee
where we have assumed a typical $N(E)\propto E^{-2}$ spectrum
generated in a mildly relativistic shock.

Shock-accelerated protons are expected to be confined in the GRB
fireball by the magnetic field.  Particles can, however, escape
directly from the internal shock region when protons convert to
neutrons through $p\gamma \to n \pi^+$ interaction.  The rate of
$p\gamma$ scattering with observed $\gamma$-rays, assumed to be
isotropically distributed in the GRB fireball, is given by
\be
K_{p\gamma} (\gamma'_p) = \frac{c}{2\gamma_p^{'2}} \int_{\veps'_{\rm
th}}^{\infty} d\veps'_r \veps'_r \sigma_{p\gamma} (\veps'_r)
\int_{\frac{\veps'_r}{2\gamma'_p}}^{\infty} 
d\veps' \frac{n_\gamma(\veps')}{\veps^{'2}}.
\label{pgamma_rate}
\ee
Here $\veps'_r = \gamma'_p \veps' (1-\beta_p \cos\theta)$ is the
photon energy evaluated in the proton's rest frame for the angle
$\theta$ between the directions of the energetic proton and target
photon, and $\veps'_{\rm th} = m_\pi c^2+ m_\pi^2c^2/2m_p$ is the
threshold photon energy for pion production.  The dominant neutron
production channel is $p \gamma \to n \pi^+$ with an intermediate
$\Delta (1232)$ baryonic resonance production.  The cross section
formula may be written in the Breit-Wigner form~\citep{sophia} as
$\sigma_{\Delta} (\veps'_r) = \sigma_0\Gamma_\Delta^2 (s/\veps'_r)^{2}
[\Gamma_\Delta^2 s + (s -m_\Delta^2)^2 ]^{-1}$, where $s=m_p^2c^4+2
\veps'_r m_pc^2$, $\sigma_0 = 3.11\times 10^{-29}$~cm$^2$ and the peak
cross section is given by $\sigma_{\rm pk}= 4.12\times
10^{-28}$~cm$^2$ at $\veps'_{r,\rm pk} = 0.3$~GeV.  The resonance
width is $\Gamma_\Delta = 0.11$~GeV.

For the photon spectrum in equation~(\ref{soft_photon_spectrum}), and
taking an approximate cross section given by
$\sigma_{p\gamma}(\veps'_r) \cong \sigma_{\rm pk}$ for $\veps'_{r,\rm
  pk} \leq \veps'_r \leq \veps'_{r,\rm pk}+\Gamma_\Delta$, the
scattering rate in equation~(\ref{pgamma_rate}) simplifies to
\ba
K (\gamma'_p) && \cong c\sigma_{\rm pk} n'_{\gamma,\rm pk}
\nonumber \\  && \times
\cases{ \frac{2^\beta (\gamma'_p \veps'_{\rm pk})^{\beta-1}}
{(\beta^2 -1) \veps_{r,\rm pk}^{'\beta-1}} - 
\frac{\veps_{\rm pk}^{'\beta-1} \veps_{r,\rm pk}^{'2}} 
{4\gamma_p^{'^2} \veps_{\rm max}^{'\beta+1} (\beta+1)} 
;~ \gamma'_p \leq \frac{\veps'_\Delta} {\veps'_{\rm pk}} \cr
{\rm ln} \left( \frac{\veps'_{r,\rm pk}} {\veps'_{\rm th}} \right)
- \frac{\veps_{r,\rm pk}^{'2}} {8\gamma_p^{'2} \veps_{\rm pk}^{'2}} 
;~\frac{\veps'_\Delta} {\veps'_{\rm sa}} > \gamma'_p \geq 
\frac{\veps'_\Delta} {\veps'_{\rm pk}} \cr
\frac{\veps_{r,\rm pk}^{'2}} {2\gamma_p^{'2} \veps_{\rm sa}^{'2}} - 
\frac{2^{1/2}\veps_{r,\rm pk}^{'5/2}} 
{5\gamma_p^{'5/2} \veps_{\rm sa}^{'5/2}}
;~\gamma'_p >
\frac{\veps'_\Delta} {\veps'_{\rm sa}}\;,}
\label{pgamma_rate2}
\ea
for $\alpha =1$.  Here $\veps'_\Delta = (\veps'_{r,\rm pk}
-\Gamma_\Delta) m_pc^2$.  The rate $ K (\gamma'_p) $ times the
dynamical time $t^\prime_{\rm dyn} \simeq r_{\rm sh}/2c\Gamma_b$
represents the opacity $\tau_{p\gamma}(\gamma'_p)$ for $p\gamma$
scattering.  For target photons with energies below
$\varepsilon^\prime_{pk}$, the $p\gamma$ scattering rate of
high-energy protons is almost constant with energy when $\alpha = 1$.
Therefore $\tau_{p\gamma}(\gamma'_p) \approx K(\gamma'_p)
t^\prime_{\rm dyn} \sim 1$ in this plateau region of $\gamma'_p$,
where $ K (\gamma'_p)$ refers to the accurate rate, equation
(\ref{pgamma_rate}), or the approximate rate, equation
(\ref{pgamma_rate2}), respectively.

The flux $J_n (E_n)$ of neutrons from $p\gamma$ interactions depends
crucially on the shocked-fireball radius, as discussed above. Assuming
that they don't interact further in the fireball,
\be
J_n (E_n) \cong 
J_p (E_n/y) t'_{\rm dyn} K(E_n/y\Gamma_b)/2y\;,
\label{neutron_flux}
\ee
where $y=E_n/E_p \approx 0.8$ is the fraction of proton energy given
to a secondary neutron, where a mean inelasticity is used.  The
injected primary proton flux (topmost thick dashed line) and escaping
neutron flux are plotted for the full numerical calculation (thin
solid line overlayed on dots) and for the approximate expression
(dots) in Fig.~\ref{fig:fluxes}.

\section{Cosmic ray interaction with stellar wind}

Massive stars such as GRB progenitors lose mass by blowing a
spherically symmetric and steady wind. We assume a nominal mass loss
rate of ${\dot M}_w = 10^{-4.5} M_{w,-4.5} M_\odot$~yr$^{-1}$ and a
wind velocity of $v_w = 10^{8} v_{w,8}$~cm~s$^{-1}$.  The volume
density of particles in the wind is ${\dot M}_w /(4\pi r^2 v_w m_p)$
and the column density at a radius $r = r_{\rm sh} = 10^{12}r_{12}$~cm
is
\be
\Sigma_w \approx 9.5\times 10^{23}
{\dot M}_{w,-4.5} v_{w,8}^{-1} r_{12}^{-1}~{\rm cm}^{-2}.
\label{wind_column_density}
\ee
The stellar wind may have high magnetic field, as has been suggested
by many authors \citep[e.g.,][]{vb88,bc93}. We assume for
simplicity that this field is in equipartition with the wind kinetic
luminosity ${\dot M}_w v_w^2/2$~\citep{wrmd07}, so that
\be
B_w \approx 141 ~w_{B,-1}^{1/2} {\dot M}_{w,-4.5}^{1/2}
v_{w,8}^{1/2} r_{12}^{-1} ~{\rm G} ~,
\label{wind_B_field}
\ee
where $w_B = 0.1w_{B,-1}$ is the equipartition parameter.
 
Cosmic ray neutrons escaping from the GRB internal shocks can interact
with dense stellar wind particles and produce secondary pions, kaons,
and higher-order resonances through $pn$ interactions.  The neutron
decay radius is $\gg c\tau_\beta \Gamma_b \approx
10^{16}\Gamma_{b,2.5}$~cm, so that neutrons that do interact with wind
particles do so before they decay, and will therefore make beamed
secondaries that would be directed along the GRB jet.  Neutral pion
and eta mesons decay almost instantaneously to produce ultrahigh
energy $\gamma$-rays.

The $\gamma$-ray flux from $pn$ interactions of neutrons with
stellar wind can be calculated from the expression
\be
J_\gamma (E_\gamma) = \Sigma_w \int_0^1 \frac{dx}{x} 
J_n \left[ \frac{E_\gamma}{x} \right] 
\sigma_{pp} \left[ \frac{E_\gamma}{x} \right] 
Y_\gamma (x; E_\gamma).
\label{gamma_emission_high} 
\ee 
Here $\sigma_{pp}(E_p)$ is the inelastic $pp$ cross-section,
$x=E_\gamma /E_n$ is the fractional $\gamma$-ray energy and $Y_\gamma
(x; E_\gamma)$ is the $\gamma$-ray yield function from neutral meson
decays.  We use the $Y_\gamma (x; E_\gamma)$ as recently
parametrized by~\citet{kab06} of the SIBYLL code which include
$\gamma$-ray production from both $\pi^0$ and $\eta^0$ decays.
Note that the charged lepton flux from pion decays may also be
calculated using equation~(\ref{gamma_emission_high}) with a change of
subscript $\gamma \to e$ and using the appropriate yield function.
The $\gamma$-ray (thin dash-dotted line) and electron (thin dashed
line) source fluxes are plotted in Fig.~\ref{fig:fluxes}

Comparing the synchrotron cooling time scale $t_{e,\rm syn} =
(3/2)\hbar^2 (B_\perp/B_Q)^{-2} (r_e m_e c E_e)^{-1}$ for $\pi^\pm$
decay $e^\pm$ in the wind magnetic field given by
equation~(\ref{wind_B_field}), with the observed $\gamma$-ray
variability time $t_{v} \simeq r_{\rm sh}/2\Gamma_b^2c$, we find that
electrons with $E_{e,\rm min} \gtrsim 8\times 10^4$~GeV radiate away a
large fraction of their energy.  Here $r_e$ is the classical electron
radius, and we let the perpendicular magnetic field $B_\perp \cong
B_w$.  The total synchrotron power emitted by an electron is given by
$P = (2/3)(r_e/\hbar^2)(B_\perp/B_Q)^2 E_e^2 m_e c$ with a
characteristic photon energy $E_c = (3/2) (B_\perp/B_Q) E_e^2/m_ec^2$,
similar to the expression in equation~(\ref{nu_m}). To a good
approximation we can assume that the total power $4\pi d_L^2 t_v P E_e
J_e(E_e)$ is emitted in photons of energy $E_c$.  The corresponding
synchrotron flux by the electrons is therefore given by
\be
E_\gamma^2 J_{\rm syn} (E_\gamma) 
\cong \frac{t_v r_e c^4 E_\gamma^{3/2}} {(3/2m_e)^{5/2}\hbar^2} 
\left( \frac{B_w}{B_Q} \right)^{1/2}  J_e (\xi).
\label{sync_spectrum}
\ee
Here $\xi = [(2/3)(B_w/B_Q)E_\gamma m_ec^2]^{1/2}$.  Note that the
synchrotron energy loss formula assumed here applies in the classical
limit defined by the parameter $\chi = (3/2) (B_\perp/B_Q)
(E_e/m_ec^2) \ll 1$.  Thus the minimum and maximum synchrotron photon
energies for the parameters adopted here may be calculated as
$E_{\gamma,\rm min} \simeq (27/2) (B_Q/B_w)^3 \hbar^4 \Gamma_b^4/
[m_e^3 c^2 r_e^2 r_{\rm sh}^2] \approx 60$ GeV and $E_{\gamma,\rm max}
\simeq (2/3) (B_Q/B_w) \chi^2 m_ec^2 \approx 10^4 \chi_{-2}^2$ GeV,
assuming $\chi = 10^{-2}\chi_{-2}$. The synchrotron flux is plotted in
Fig.~\ref{fig:fluxes} with an exponential cutoff above $E_{\gamma,\rm
  max}$.

Compton losses on the scattered stellar radiation field can be shown
to be small compared with synchrotron losses.  The energy density of
scattered stellar photons from the pre-burst star is $\approx L_*
\tau_w/4\pi r^2 c \cong 260~L_{*,38}\tau_w/r^2_{12}$ ergs cm$^{-3}$,
where $L_* = 10^{38} L_{*,38} {\rm~ergs~s}^{-1}$ is the pre-burst
stellar luminosity and $\tau_w$ is the Thomson depth of the wind.
This is smaller than the magnetic field energy density $B_w^2/8\pi$
given from the expression for $B_w$ in equation (\ref{wind_B_field}),
even for a luminous pre-burst star.  Klein-Nishina effects will make
the Compton losses even smaller.  TeV $\gamma$ rays might also produce
$e^\pm$ pairs with the stellar photons through $\gamma\gamma$
interactions.  The optical depth of TeV photons to $\gamma\gamma$ pair
production with stellar photons with mean energy $\bar \epsilon_*$ can
be written as $\tau_{\gamma\gamma} \simeq (\sigma_{\rm T}/3)
n_{ph}(\bar \epsilon_* ) r \simeq 4\times 10^{-5} L_{*,38} \tau_{\rm
  T}/[r_{12} (\bar \epsilon_*/{\rm eV})]$.  As can be seen, this
process can be neglected.

High energy $\gamma$-rays are also subject to absorption with photons
of the EBL while propagating from the source to Earth.  The opacity
$\tau_{\gamma\gamma} \sim 1$ for $\approx 600$~GeV photons from a
source at $z\approx 0.1$~\citep{rdf08}.  To calculate the opacity, we
assumed that the background radiation field consists of three
components---cosmic microwave background, infrared and optical
photons---represented by a blackbody spectrum, a modified blackbody
spectrum~\citep{dermer07} and a fit~\citep{rdf08}, respectively.  The
final emerging $\gamma$-ray spectrum (thick solid curve) is plotted in
Fig.~\ref{fig:fluxes}.  The absorbed $\gamma$-rays can induce a pair
cascade and give rise to a long duration component after the burst is
over if the intergalactic magnetic field is sufficiently weak,
$\lesssim 10^{-16}$ G~\citep{rmz04}.

\section{Results and Discussion}

\begin{figure} \plotone{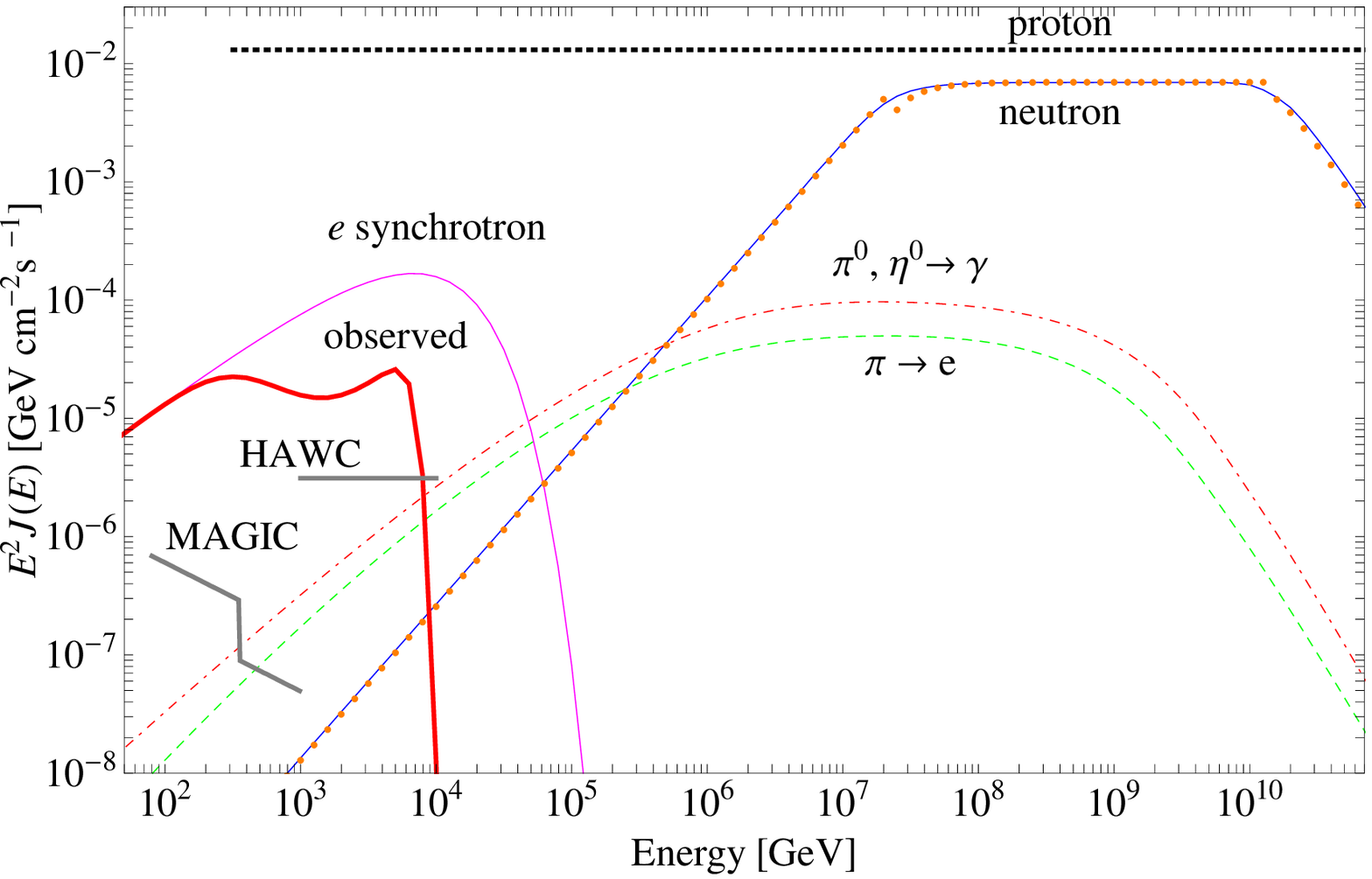} \caption{Spectra of shock-accelerated
cosmic rays, neutrons escaping from the GRB fireball (thin solid line
shows the numerical calculation using the full $p\gamma \to \Delta$
cross section formula, and the dots show equation (\ref{pgamma_rate2})
using a constant $p\gamma$ cross section), $\pi^0$ and $\eta^0$ decay
$\gamma$-rays and $\pi^\pm$ decay electrons.  The $e$ synchrotron
radiation is calculated from pion-decay electrons in the wind magnetic
field, and the thick solid line is the emerging $\gamma$-ray spectrum
after absorption in background radiation fields. Also plotted are the
MAGIC and HAWC detector sensitivities.}  \label{fig:fluxes}
\end{figure}

Figure~\ref{fig:fluxes} shows the results of our study.  We assumed a
GRB luminosity distance of $d_L = 455$~Mpc ($z=0.1$), with all scaling
parameters equal to unity (most importantly, $r_{12} = 1$ and
$M_{w,-4.5} = 1$).  The ``observed" $\gamma$-ray spectrum is
calculated from the ``$e$ synchrotron" and ``$\pi^0,~\eta^0 \to
\gamma$" components after taking into account absorption in the EBL.
Also shown in Fig.~\ref{fig:fluxes} are the detection sensitivites of
the MAGIC \citep{magic06,sca06} and HAWC (see footnote 4)
detectors. For MAGIC, we used their 60 s $5\sigma$ GRB sensitivity of
5.8~Crab between 80~GeV -- 350~GeV and 1.8~Crab between 350~GeV --
1~TeV.  For HAWC we used their 10~s $5\sigma$ GRB sensitivity within
0--10 degrees of the azimuth.

As shown in Fig.~\ref{fig:fluxes}, a typical long duration GRB inside
a stellar wind environment may be detected by IACTs with rapid slewing
capability such as MAGIC, VERITAS or HESS and by the upcoming HAWC
detector.  If all long duration GRBs within $z\sim 0.1$ have dense
stellar wind as modeled here, then the expected detection rate in
upcoming TeV detector would be $\gtrsim 1$~burst~yr$^{-1}$, using a
GRB rate of $2$~Gpc$^{-3}$~yr$^{-1}$.  HAWC would be sensitive for
GRBs with $z \lesssim 0.2$, when distance and EBL effects become
important.

In the context of the internal shock model, a prompt TeV emission
signal will result from the first pair of colliding shells that form
neutrons which escape and then interact with particles in the wind. As
the merged shell moves out along the GRB jet, it sweeps up wind
material, so that subsequent escaping neutrons will no longer have
target wind particles with which to interact.  The column density of
material does not however change, so that neutrons formed by further
pairs of colliding shells still have a significant target column
density with which to interact and make TeV radiation.  Thus the
duration of the prompt TeV signal in this model corresponds to the
duration of the prompt phase associated with colliding shells.

Predicted TeV $\gamma$-ray emission in the early afterglow phase,
either by hadronic interactions~\citep{bd98} or synchrotron
self-Compton (SSC) emission~\citep{dermer00,wdl01,zm01}, is expected
to last much longer than the prompt TeV emission considered here. A
leptonic SSC origin of TeV radiation formed by an external shock will
correlate with the lower energy synchrotron radiation with a peak
energy that becomes smaller as the blast wave decelerates.  By
contrast, the TeV emission formed by the process considered here will
end when the central engine becomes inactive.  The TeV $\gamma$-ray
flux predicted in this work should correlate with the activity of the
central engine as reflected by the MeV emission from a GRB, though
delays could arise from the time required to accelerate protons to
ultra-high energies.  High-energies neutrinos are formed directly from
photopion-producing interactions in the internal shocks (in the
TeV--PeV energy range), as well as from $pn$ interactions (in the
PeV--EeV energy range) in the stellar wind.  Joint detection of prompt
high-energy neutrinos and prompt TeV radiation would provide a new
method to probe the environment in the vicinity of GRBs.

\acknowledgements 
The work of S.R and C.D.D. is supported by the
Office of Naval Research.

\clearpage 

\end{document}